\documentclass{elsarticle}
\usepackage{amsmath,amssymb,a4wide,psfrag,graphicx,hyperref}

\def\Dslash{\mathchoice
    {D\hskip-0.62em\raise0.2ex\hbox{$\displaystyle/$}\hskip0.2em}%
    {D\hskip-0.62em\raise0.2ex\hbox{$\textstyle/$}\hskip0.2em}%
    {D\hskip-0.5em\raise0.15ex\hbox{$\scriptstyle/$}\hskip0.2em}%
    {D\hskip-0.5em\raise0.15ex\hbox{$\scriptscriptstyle/$}\hskip0.2em}}
\def\dslash{\mathchoice
    {\partial\hskip-0.5em\raise0.2ex\hbox{$\displaystyle/$}\hskip0.2em}%
    {\partial\hskip-0.5em\raise0.2ex\hbox{$\textstyle/$}\hskip0.2em}%
    {\partial\hskip-0.4em\raise0.15ex\hbox{$\scriptstyle/$}\hskip0.2em}%
    {\partial\hskip-0.4em\raise0.15ex\hbox{$\scriptscriptstyle/$}\hskip0.2em}}
\def\Aslash{\mathchoice
    {A\hskip-0.5em\raise0.2ex\hbox{$\displaystyle/$}\hskip0.2em}%
    {A\hskip-0.5em\raise0.2ex\hbox{$\textstyle/$}\hskip0.2em}%
    {A\hskip-0.4em\raise0.15ex\hbox{$\scriptstyle/$}\hskip0.2em}%
    {A\hskip-0.4em\raise0.15ex\hbox{$\scriptscriptstyle/$}\hskip0.2em}}

\newcommand{\Tr}[0]{\text{Tr}}
\newcommand{\sign}[0]{\text{sign}}

\newcommand{\ket}[1]{\left|#1 \right\rangle}
\newcommand{\bra}[1]{\left\langle#1 \right|}

\renewcommand{\qed}{\nobreak \ifvmode \relax \else
      \ifdim\lastskip<1.5em \hskip-\lastskip
      \hskip1.5em plus0em minus0.5em \fi \nobreak
      \vrule height0.75em width0.5em depth0.25em\fi}

\begin{document}
\begin{frontmatter}
\title{Modifying the molecular dynamics action to increase topological tunnelling rate for dynamical overlap fermions}
\author[a]{Nigel Cundy}
\author[a]{Weonjong Lee}
\address[a]{Lattice Gauge Theory Research Center, FPRD, and CTP,\\ Department of Physics \&
    Astronomy, Seoul National University, Seoul, 151-747,\\ South Korea}
\date{\today}
\begin{abstract}
We describe a new Hybrid Monte Carlo (HMC) algorithm for dynamical overlap
fermions, which improves the rate of topological index changes by adding an
additional (intensive) term to the action for the molecular dynamics part of the
algorithm. The metropolis step still uses the exact action, so that
the Monte Carlo algorithm still generates the correct ensemble. By tuning this
new term, we hope to be able to balance the acceptance rate of the HMC algorithm
and the rate of topological index changes. We also describe how suppressing, but
not eliminating, the small eigenvalues of the kernel operator may improve the
volume scaling of the cost per trajectory for overlap HMC while still allowing
topological index changes.

We test this operator on small lattices, comparing our new algorithm with an old overlap HMC algorithm with a slower rate of topological charge changes, and an overlap HMC algorithm which fixes the topology. Our new HMC algorithm more than doubles the rate of topological index changes compared to the previous state of the art, while maintaining the same metropolis acceptance rate. We investigate the effect of topological index changes on the local topological charge density, measured using an improved field theoretic operator after heavy smearing. We find that the creation and annihilation of large lumps of topological charge is increased with the new algorithm.
\end{abstract}
\end{frontmatter}
\section{Introduction}\label{sec:1}
Recently, there has been a discussion concerning the autocorrelation of topological observables for Wilson and quenched fermions in the continuum limit~\cite{Schaefer:2010hu}. It has been found that as the continuum limit is approached, it becomes increasingly hard for the Hybrid Monte-Carlo routine to tunnel between topological sectors. The transfer matrix for the Markov chain has a disperse spectrum, with the smallest eigenvalues having a strong dependence on the lattice spacing. The consequences of this autocorrelation for a particular observable depend on the extent to which the observable couples to the slow mode of the Markov chain. It was found that the global topological charge has a particularly strong coupling to these particularly slow modes; but it may well be that there is an effect, albeit small, for many other observables. It is difficult to know \textit{a priori} which observables couple strongly to which eigenvectors of the transfer matrix, although quantities related to the topology of the 
QCD vacuum seem to be particularly strongly effected. 

What are the effects of such a long autocorrelation? Firstly, in the best
scenario, an underestimation of the statistical error. Secondly, it could lead
to an effective breakdown in ergodicity\footnote{i.e. the time required to
correctly sample different sectors of the configuration space is larger than
what is practical.} and therefore an incorrect result. It is not clear that each
topological sector is internally connected at small lattice spacing, that is to
say that it is in practice possible to progress from a 1+1 configuration (one
instanton and one anti-instantons, or the equivalent in monopoles, vortices,
membranes or whatever model of the QCD vacuum the reader prefers) to a 2+2
configuration without progressing through a 2+1 or 1+2 configuration; nor is it
clear that the phenomenology for each of these configurations is the same for
every observable. At coarse lattice spacing, one can argue that an
instanton/anti-instanton pair may slip through the gaps in the lattice; at fine
lattice spacing this will become harder.
 Even if the topological sectors are internally connected, the autocorrelation
may be large enough in comparison to the total Monte-Carlo time of a typical run
that it fails to sample every sector. Nor is it clear that the physics of
these ensembles is the same. This is particularly important when
considering the thermalisation of the ensemble: the configurations may not even
reach the correct average density of topological fluctuations, and it will be
difficult to determine this just from the ensembles.


Of course, this comes as little surprise. In the continuum, the topological sectors are disconnected (one cannot map from one topological sector to another with a continuous transformation of the gauge fields or, equivalently, the Dirac operator). As the continuum limit is approached, as long as the theory is in the right universality class, we should expect difficulties with changing topological charge. One can indeed say that if an action does not have difficulties concerning the topological autocorrelation at some point as the continuum limit is approached then there is a distinct possibility that the action is not in the same universality class as the continuum.

While the auto-correlation of the global topological charge is not expected to be significant in the infinite volume limit, the density and size of local fluctuations of the charge density are likely to affect many observables (as seen, for example, in instanton models of the vacuum~\cite{Instantons2}). An important question is therefore to what extent does the autocorrelation of local topological fluctuations depend on the rate of global topological charge changes. 

Topological autocorrelation is also a problem for Ginsparg-Wilson and approximate Ginsparg-Wilson Dirac operators at any lattice spacing. These have the advantage of possessing a well defined topological index; and resemble the continuum theory in that the topological sectors are disconnected: again, one cannot map from one topological sector to another with a continuous transformation of the Dirac operator. In this sense, they are far better equipped than the Wilson or staggered actions and their descendants to analyse the problems of topological autocorrelation. This is clearly seen for the overlap operator~\cite{Narayanan:1993sk,Narayanan:1993ss,Neuberger:1998fp,Neuberger:1997bg,Neuberger:1998my},
\begin{gather}
D[\mu] = 1+\mu + \gamma_5 (1-\mu) \sign(K) 
\end{gather}
where 
\begin{gather}
K = \gamma_5 (D_W - m),
\end{gather}
and (for example) we may use a smeared and perhaps improved Wilson operator for $D_W$, and the bare quark mass is $2\mu/((1-\mu)m)$, with $0\lesssim m<2$. Once the action is fixed by choosing $D_W$ and $m$ and using an overlap representation of the Yang-Mills term, for example,
\begin{gather}
S = \beta \Tr\left(D^{\dagger}[0] D[0] - \frac{1}{4}\Tr_s D^{\dagger}[0] D[0]\right)^2 + \sum_i \overline{\psi}_i D[\mu_i]\psi_i,\label{eq:1} %
\end{gather}
where $\Tr_s$ indicates the trace of the spinor indices only (multiplied by the identity matrix), and $\beta$ is inversely proportional to the square of the bare coupling, then the topological index is uniquely defined for that ensemble. Constructing the Yang Mills action from the Dirac operator ensures a consistent definition of the topological index in both sectors of the action~\cite{Liu:2007hq,Horvath:2006aj,Horvath:2006md}, and this particular construction~\cite{Cundy:2009ae} is chosen because the Lagrangian is clearly local and it explicitly gives the correct result for the continuum Dirac operator. However, this Yang Mills action is expensive to simulate for lattice chiral fermions (even more than for the fermion determinant itself), so for the moment we are restricted to more traditional gauge actions, such as the plaquette, Symanzik improved~\cite{SI}, or, as used in this study, a tadpole improved Luscher Weisz gauge action~\cite{TILW,TILW2,TILW3,TILW4}. 

This Dirac operator has a topological index
\begin{gather}
Q = - \frac{1}{2} \Tr\; \sign(K),
\end{gather}
which is the difference between the number of positive and negative eigenvalues of $K$. In the continuum limit, this index is equal to the global topological charge (shown for overlap fermions in~\cite{Adams:2000rn} and fixed point fermions in~\cite{Hasenfratz:1998ri}). The fermion determinant, after integrating out the fermion fields in the action, has a discontinuity when there is a change in the topological index, which occurs when the eigenvalue $\lambda_0$ associated with the eigenvector $\psi_0$ crosses zero as the gauge field is smoothly modified:
\begin{align}
\Delta S_{\textit{real}} =& \Tr\log\left(\frac{1}{D}(D - 2(1-\mu)\gamma_5 \sign(\lambda_0) \ket{\psi_0}\bra{\psi_0})\right)\nonumber\\
=& \Tr\log\left(1-(1-\mu)\frac{2\sign(\lambda_0)}{\gamma_5 D} \ket{\psi_0}\bra{\psi_0}\right)
\nonumber\\ = & \log \left(1 - (1-\mu)\bra{\psi_0}\frac{2\sign(\lambda_0)}{\gamma_5 D} \ket{\psi_0}\right),\label{eq:deltaSreal}
\end{align}
where the Dirac operator $D$ and $\lambda_0$ are calculated just before the eigenvalue crosses zero.
Similarly, if the overlap definition of the Yang-Mills action is used, as in
equation (\ref{eq:1}), there will be a discontinuity in this part of the action
as well. A discontinuous action can be incorporated into a Hybrid Monte-Carlo
algorithm~\cite{HMC} using a transmission/reflection
routine~\cite{Fodor:2003bh,Cundy:2005pi, Cundy:2005mr}, where the eigenvector
may either reflect of the potential wall caused by this discontinuity in the
action if the contribution to the kinetic energy from the momentum perpendicular
to the barrier is smaller than the action discontinuity, or transmit through the
barrier with reduced momentum if the kinetic energy is larger. In the first
case, there is no topological index change; in the second there is. For a
Gaussian momentum distribution, the probability of transmission for an action
discontinuity $\Delta S$ is (approximately, with the exact expression depending
on which of the various variants of the transmission/reflection algorithm is
used) $\min(1,e^{-\Delta S})$~\cite{
Cundy:2005pi}, and if the topological autocorrelation depends on the rate of topological index changes then the distribution of $\Delta S$ is crucial in determining the autocorrelation. Additionally, the more flavours one adds the larger $\Delta S$ will be and the harder it will be to tunnel between topological sectors. This action discontinuity has only a weak dependence on the quark mass~\cite{Cundy:2007la,Cundy:2008zc}.

For overlap actions, the rate of topological index changes may be suppressed by at least one of two factors:
\begin{itemize}
\item A large $\Delta S$, leading to infrequent transmissions when there is an attempted eigenvalue crossing.
\item A low density of kernel eigenvalues around zero, leading to a low rate of attempted eigenvalue crossings.
\end{itemize}
For continuous actions (including in the Yang-Mills part of the action), only the second of these can occur (the topological index may still be calculated using an overlap operator with a particular kernel; up to lattice artefacts in most normal situations this will be the same as the natural definition of the index for the chosen action), so at small lattice spacing there has to be a suppression of kernel eigenvalues. 
We expect $\Delta S$ to increase as the lattice spacing decreases.  

For overlap actions, the suppression of small kernel eigenvalues is useful in the sense that it speeds up the simulation, which is heavily dependent on the condition number of the kernel operator and on the number of attempted eigenvalue crossings.  If the topological auto-correlation reduces as there are more topological charge changes, then the key to the performance of the overlap operator is the probability of transmission, and one of the efforts of recent work on overlap Monte-Carlo algorithms, is to reduce $\Delta S$ as much as possible (see \cite{Cundy:2007la,Cundy:2008zc} and section \ref{sec:2.3} below). The noisy pseudo-fermion estimate of the determinant gives a far worse $\Delta S$ than would be seen in the exact theory; by factorising the determinant and treating the part for small kernel eigenvalues precisely one is able to remove this noise and reduce $\Delta S$ to what would be seen for the exact determinant. This is the best that can be achieved for an energy conserving HMC algorithm. It is 
here that we hope that overlap fermions (with an overlap gauge action) may gain
over other types of fermion. Because the discontinuity in the action is
explicit, it may be more easily treatable. Due to universality, as the continuum
limit is approached, the same poor mass scaling of the equivalent of $\Delta S$
with a purely pseudo-fermion estimate of the determinant must make itself
visible in some way for any valid lattice action. 

However, even modifying the algorithm in this way leaves a too large $\Delta S$
for a rapid tunnelling rate for some ensembles ($\Delta S$ depends on the
lattice spacing, temperature, volume and number of fermion flavours). It seems
that, as we approach the continuum and small residual mass limits, Lattice QCD
actions with good chiral symmetry are doomed to have low rates of topological
tunnelling. One reason chiral fermions are used rather than some cheaper action
is to ensure that the topological properties of continuum QCD are treated
correctly; but if the autocorrelation is too bad, this will not be the case.

We note that the rate of topological index changes with overlap fermions depends on the HMC algorithm used. Different algorithms allow different rates of topological charge changes. For some algorithms (adding a ghost term to the action to fix the topology, or using a simple pseudo-fermion estimate of the determinant without determinant factorisation), topological charge changes are (for all practical purposes) impossible for ensembles with quark masses approaching the physical values. For others, such as determinant factorisation, the topological index changes more rapidly. Each of these algorithms generates ensembles which ought to describe precisely the same physics (baring the small corrections from the ghost terms when fixing the topology) were the gauge fields sampled correctly: the only difference is in terms of autocorrelation and ergodicity. Overlap fermions thus provide a good place to test the effects of suppression of global topological index changes.

It is therefore desirable to have an overlap HMC algorithm which samples different topological sectors as efficiently as possible. The previous work gave the best possible rate of topological index changes for a molecular dynamics which conserves the HMC energy. To beat this, we require a molecular dynamics which does not conserve the HMC energy. In this work, albeit using small lattices, large quark masses and large lattice spacings for an initial study, we compare a new overlap algorithm which improves the tunnelling rate with the old (and our current production) determinant factorised algorithm~\cite{Cundy:2007la} and with a topology fixed algorithm where small kernel eigenvalues are suppressed by an additional term in the action~\cite{Hashimoto:2006rb}). Our new algorithm does not conserve the action exactly across the topological charge barrier, instead moving some of the cost into the metropolis accept/reject step. By tuning various parameters, we hope to be able to maintain a good acceptance rate 
while increasing the rate of topological charge changes. 

Section \ref{sec:motiv} motivates why we are interested in increasing the rate
of topological index changes. Section \ref{sec:algorithm} describes the
algorithms and actions compared in this study; section \ref{sec:results} gives
our results, and section \ref{sec:conclusions} concludes. There is an appendix
giving a little more detail concerning the topological charge density operator
used in section \ref{sec:motiv}.

\section{Motivation}\label{sec:motiv}
It may be asked why we wish to increase the rate of topological charge changes.
After all, the global topological charge is not expected to play a role in any
observable in the infinite volume limit. However, that is not likely to be the
case for the local topological charge density, $\frac{1}{32\pi^2} F_{\mu\nu}
F_{\rho\sigma} \epsilon_{\mu\nu\rho\sigma}$. It is widely believed that
topological fluctuations in the local topological charge density, such as
vortices~\cite{Vortices}, monopoles~\cite{Monopoles1,Monopoles2,Calorons} or
instantons~\cite{Instantons2,Instantons1} are responsible for two of the key
non-perturbative characteristics of QCD, confinement and chiral symmetry
breaking. Each of these objects is associated with localised lumps or lines of
topological charge which in total give an integer charge. To correctly sample
configuration space involves correctly sampling the possible configurations of
these topological objects. This may be achieved either by creating and
annihilating single (anti-)
instantons (or vortices, or monopoles), which requires global topological index changes, or by creating or annihilating instanton--anti-instanton pairs, and then letting the pairs move apart and around the lattice.

\begin{figure}
\begin{center}
\begin{tabular}{c}
\includegraphics[width = 0.85\textwidth]{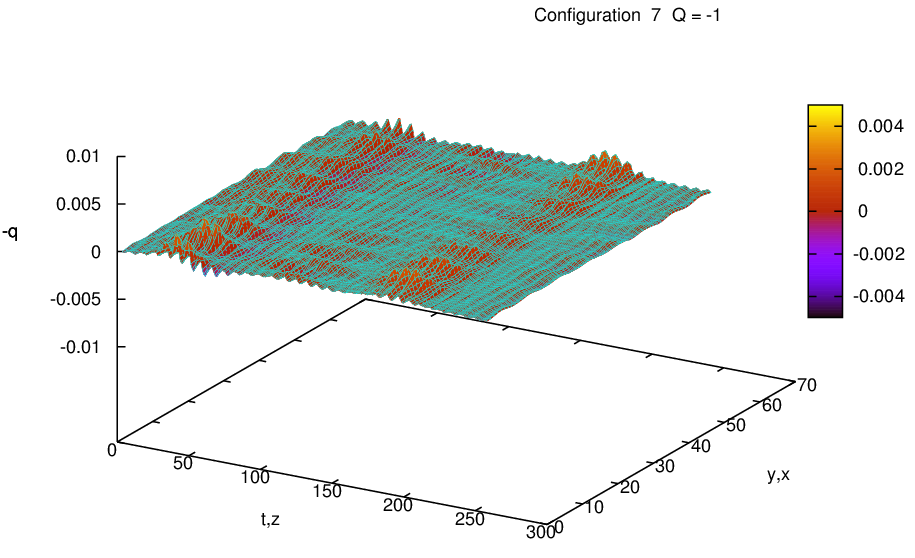}\\
\includegraphics[width = 0.85\textwidth]{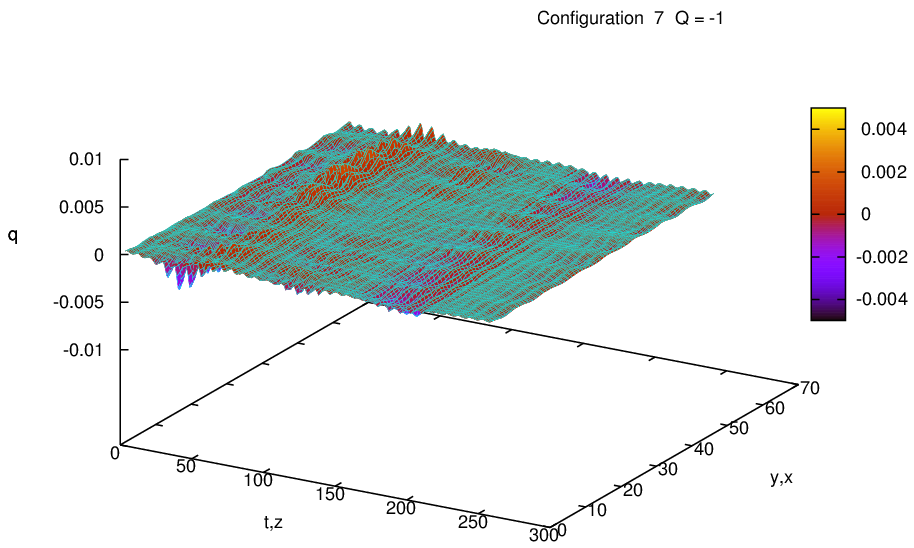}
\end{tabular}
\end{center}
\caption{The topological charge density for configuration 7 of ensemble 2 (the
top plot shows the negative charge density, the bottom plot the positive charge
density). This configuration has topological index $-1$; there are lumps of
charge $Q$ centred at ($t,x,y,z,Q$) = (24,6,8,1,-1.004), (7,1,7,4,0.600),
(5,1,1,7,-0.462) }\label{fig:fig1}
\end{figure}

\begin{figure}
\begin{center}
\begin{tabular}{c}
\includegraphics[width = 0.85\textwidth]{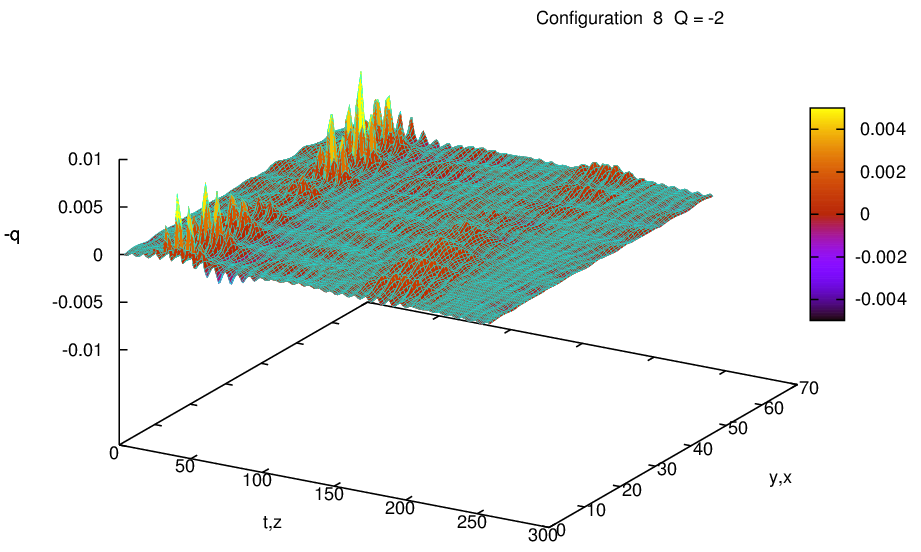}\\
\includegraphics[width = 0.85\textwidth]{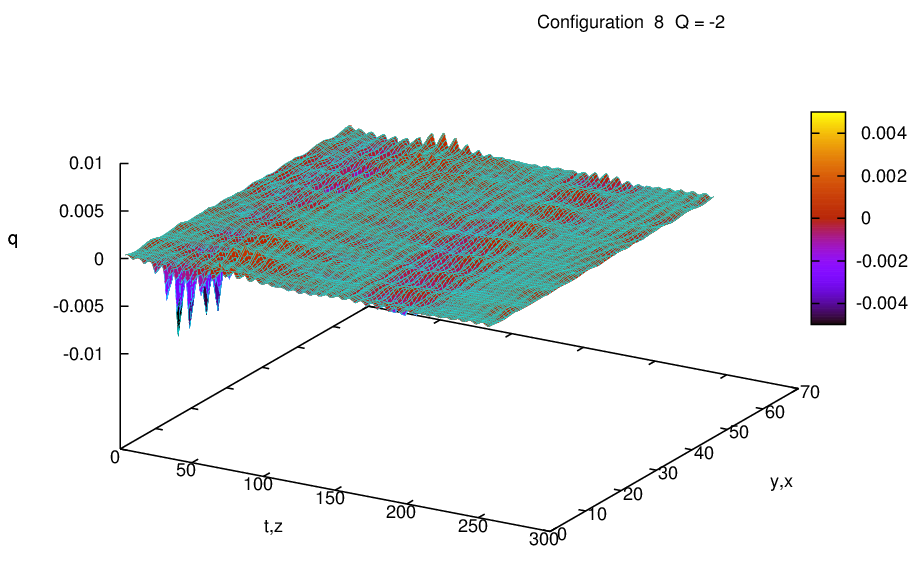}
\end{tabular}
\end{center}
\caption{The topological charge density for configuration 8 of ensemble 2 (the
top plot shows the negative charge density, the bottom plot the positive charge
density). This configuration has topological index $-2$; there are lumps of
charge $Q$ centred at ($t,x,y,z,Q$) = (5,1,1,6,-1.162),(23,6,1,8,-0.962)
}\label{fig:fig2}
\end{figure}

\begin{figure}
\begin{center}
\begin{tabular}{c}
\includegraphics[width = 0.85\textwidth]{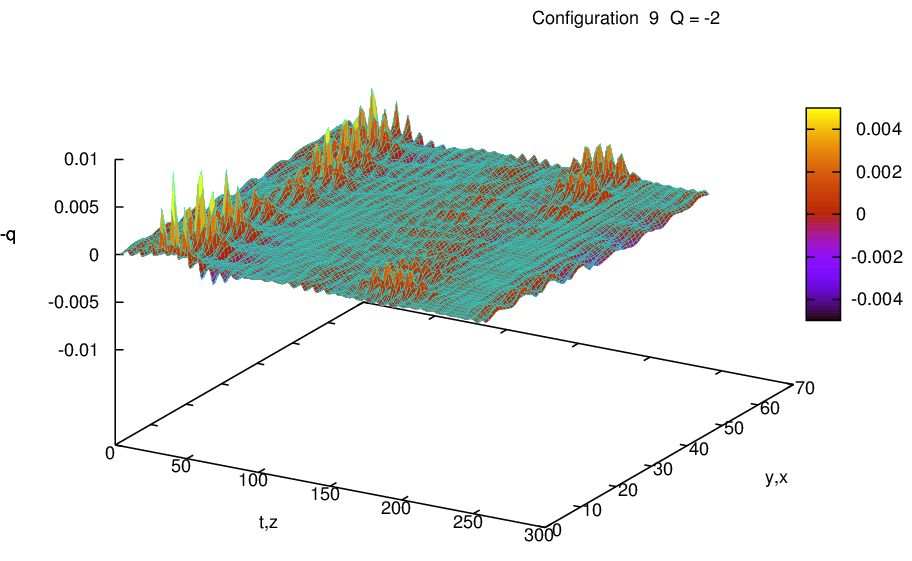}\\
\includegraphics[width = 0.85\textwidth]{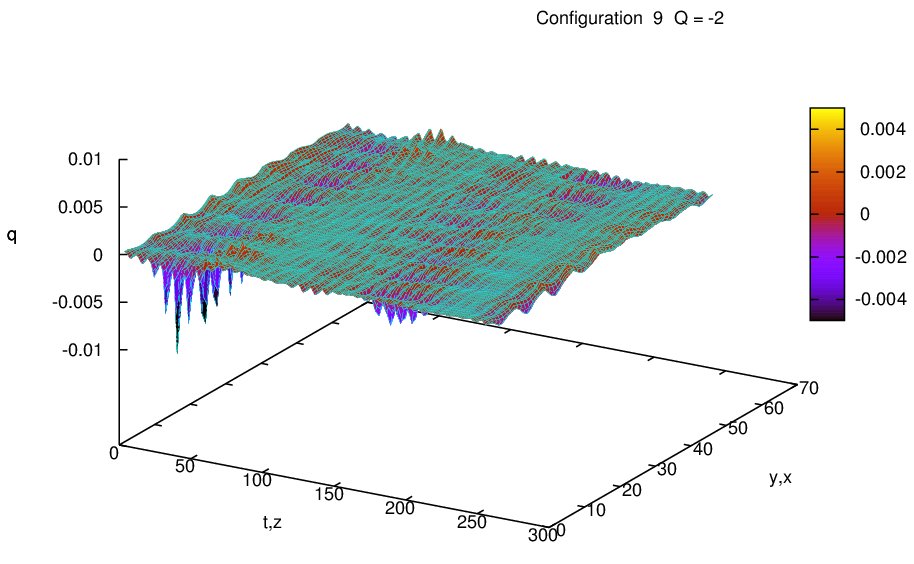}
\end{tabular}
\end{center}
\caption{The topological charge density for configuration 9 of ensemble 2 (the
top plot shows the negative charge density, the bottom plot the positive charge
density). This configuration has topological index $-2$; there are lumps of
charge $Q$ centred at ($t,x,y,z,Q$) = (5,1,1,6,-1.056), (24,5,8,8,-0.839).
}\label{fig:fig3}
\end{figure}

\begin{figure}
\begin{center}
\begin{tabular}{c}
\includegraphics[width = 0.85\textwidth]{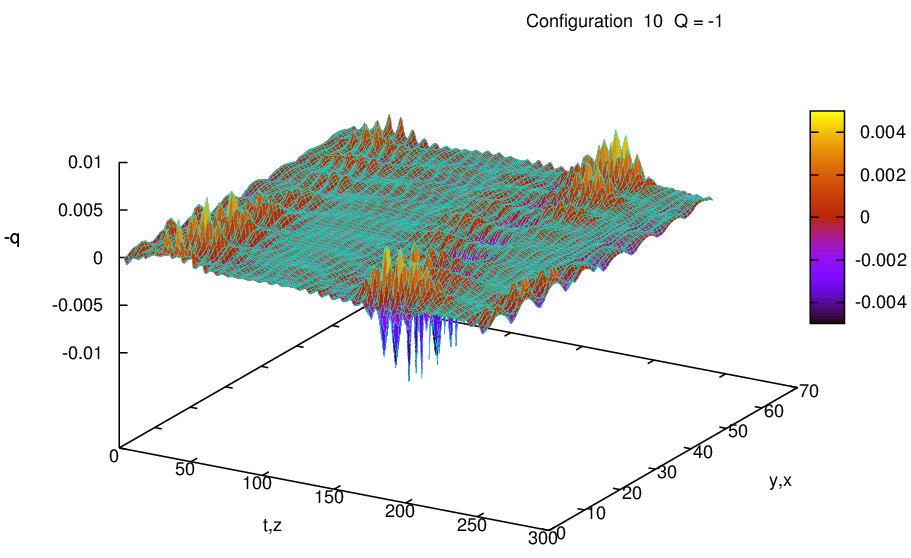}\\
\includegraphics[width = 0.85\textwidth]{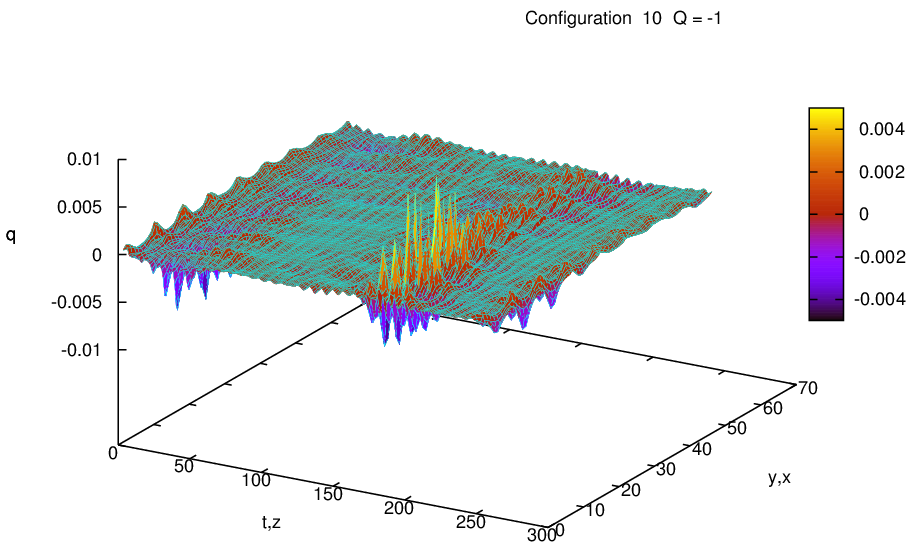}
\end{tabular}
\end{center}
\caption{The topological charge density for configuration 10 of ensemble 2 (the
top plot shows the negative charge density, the bottom plot the positive charge
density). This configuration has topological index $-1$; there are lumps of
charge $Q$ centred at ($t,x,y,z,Q$) =
(5,1,2,7,-1.233),(25,6,8,1,-1.045),(24,2,2,4,1.002)}\label{fig:fig4}
\end{figure}

The creation of instantons when there is a topological index change can be seen explicitly. In figures \ref{fig:fig1}-\ref{fig:fig4} we show the local topological charge density for configurations on one of the ensembles generated in this work (for details of the ensembles, see below; the lattice size is $8^3 32$). We used the gluonic definition of the topological charge density discussed in the appendix.

 The charge density is dominated by peaks with topological charges usually in the range $0 \sim 1$, and the occasional peak with charge $\sim 2$. The peaks appear approximately hyper-spherical. In the figures, the $x$-axis represents $L_S(t-1) + z$ and the y-axis $  L_S (y-1) + x$, where $L_S = 8$ is the spatial dimension of the lattice, and the lattice coordinates are $(t,x,y,z)$. A single peak in the charge density therefore shows up in these plots as a collection of spikes separated by $L_S$. The top plot shows the negative charge density, the bottom plot the positive charge density (the same as the top plot, but with the z axis flipped in sign). (The `lasagne' or membrane structure seen in other studies~\cite{Horvath:2003yj} emerges at smaller lattice spacing or smaller coupling, and can be seen to an extent away from the peaks; most of the absolute value of the topological charge is contained within these smaller fluctuations dispersed across the whole lattice. This charge is defined after heavy 
smearing, which distorts the gauge field by making it more semi-classical. Additional definitions of the local charge density are therefore needed as a comparison in any specific study of the QCD vacuum.) We identify the peaks as local maxima or minima of the topological charge distribution (within a hypercube  of length 3), and assign points to the lumps associated with each peak by assigning them to same lump as the lattice site within the surrounding hypercube with the highest (or lowest, for the anti-lumps) topological charge density. In this way, we can obtain a crude estimate of the size, shape, and topological charge of each of the lumps. 

On these four configurations, there are two topological index changes. The
first, between configurations 7 and 8, converts two objects of charge around
$\pm0.5$ (which we may interpret as a closely placed instanton/anti-instanton
pair into a single object of charge $\sim 1$ (which we interpret as an
instanton); the second, between configurations 9 and 10, creates an object of
charge $-1$ (which we interpret as an anti-instanton). While the lumps in these
configurations have approximately integer charge, this is not true across the
ensemble: there is a continuous spectrum of charges for the individual lumps up
to a charge of $\sim \pm 1$, with a few objects having charges $\sim\pm 2$.
Between topological index changes, these objects tend to endure, although they
move around the lattice or expand or flatten and occasionally annihilate with an
object of opposite charge. Thus a topological index charge creates a
discontinuous change in the structure of the QCD vacuum (at least, as
interpreted through the lens of heavy 
stout smearing). 

\begin{figure}
\begin{center}
\begin{tabular}{c}
\includegraphics[width = 0.85\textwidth]{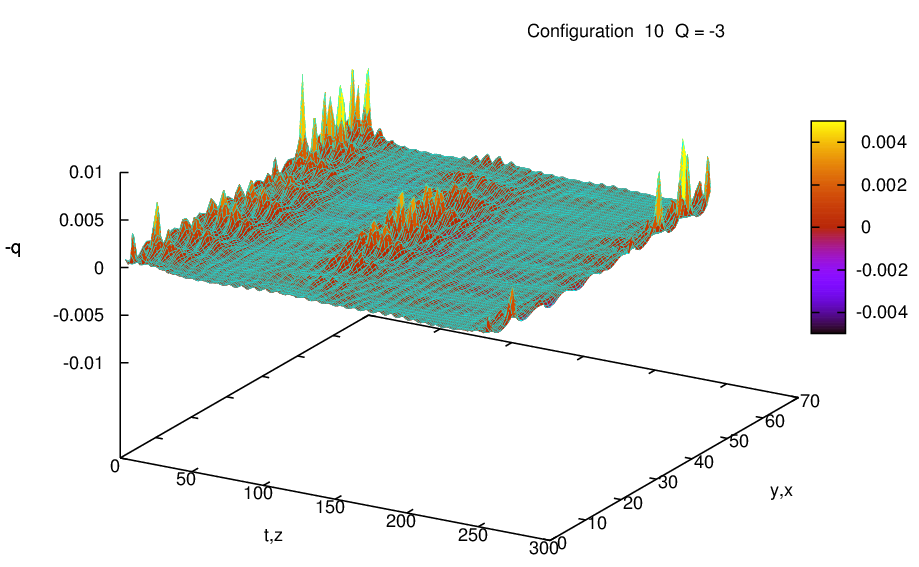}\\
\includegraphics[width = 0.85\textwidth]{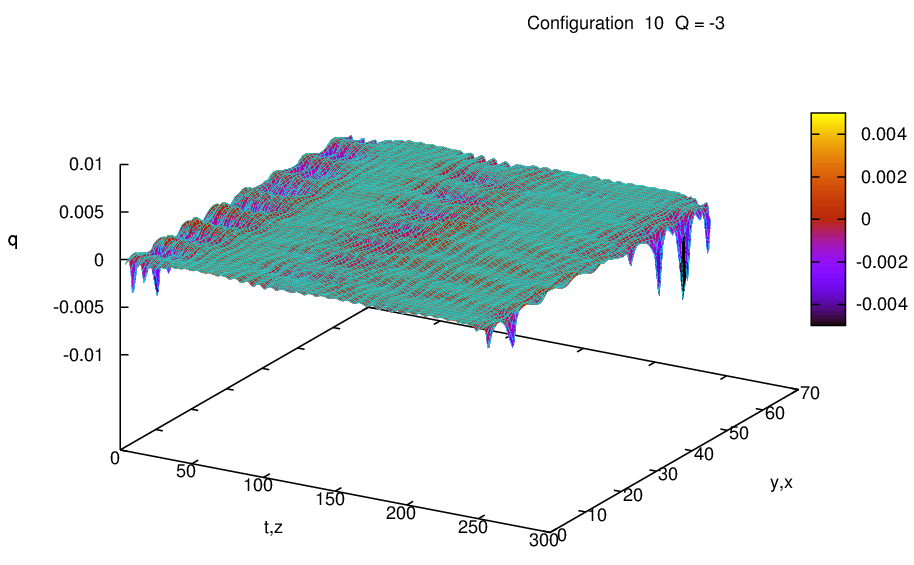}
\end{tabular}
\end{center}
\caption{The topological charge density for configuration 10 of ensemble TF1
(the top plot shows the negative charge density, the bottom plot the positive
charge density). This configuration has topological index $-3$; there are lumps
of charge $Q$ centred at ($t,x,y,z,Q$) = (1,8,8,7,-1.285), (15,8,4,6,-1.100),
(3,6,3,2,-0.783). }\label{fig:fig5}
\end{figure}

\begin{figure}
\begin{center}
\begin{tabular}{c}
\includegraphics[width = 0.85\textwidth]{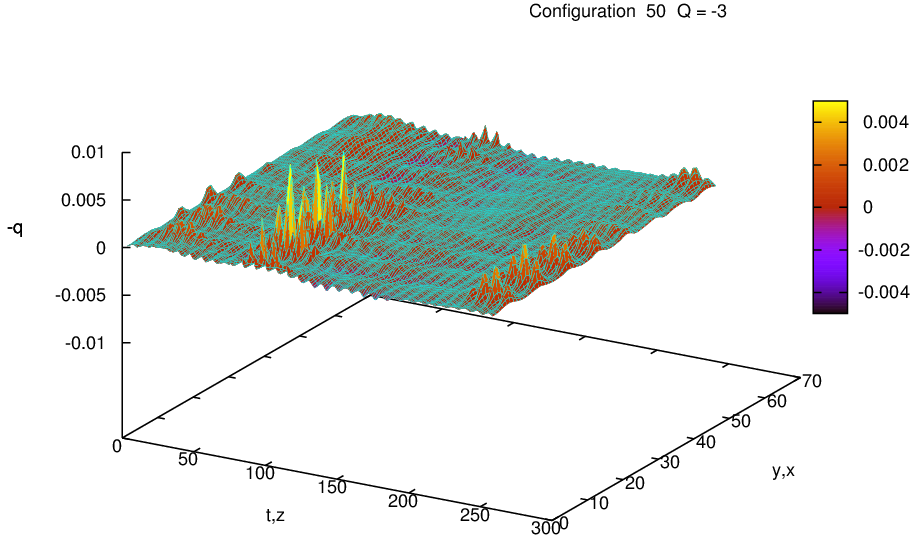}\\
\includegraphics[width = 0.85\textwidth]{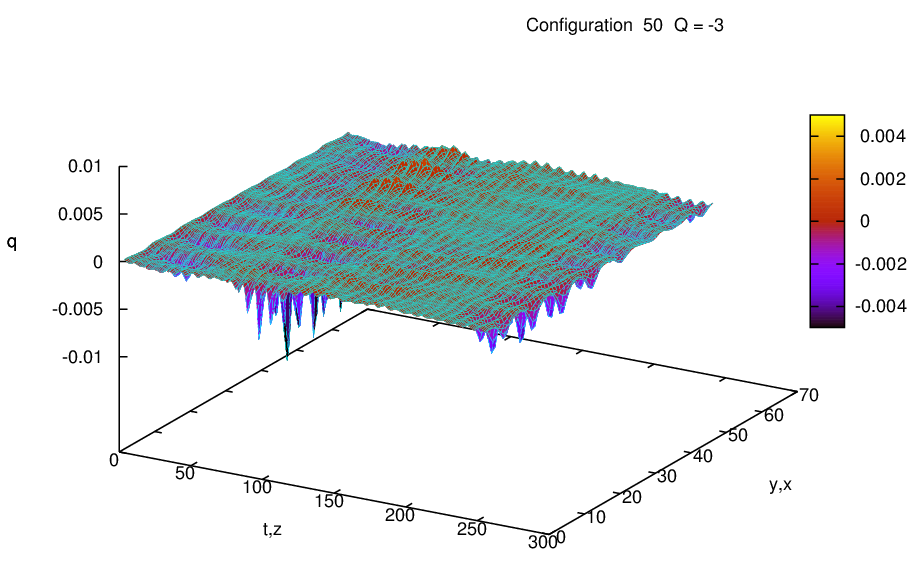}
\end{tabular}
\end{center}
\caption{The topological charge density for configuration 50 of ensemble TF1
(the top plot shows the negative charge density, the bottom plot the positive
charge density). This configuration has topological index $-3$; there are lumps
of charge $Q$ centred at ($t,x,y,z,Q$) =  (12,8,2,7,-1.070), (31,6,2,5,-0.972),
(4,6,1,8,-0.968)}\label{fig:fig6}
\end{figure}

This may be compared with what occurs at fixed large topological charge (figures \ref{fig:fig5} and \ref{fig:fig6}). This ensemble was generated with the fixed topology algorithm (see section \ref{sec:fixedtopology}) at index $-3$, which was the largest topological index seen on these lattices. Here the same structures are preserved for the entire ensemble, although they move to different lattice sites and change in size and shape. We show here configurations 10 and 50; the same three objects are discernible on each plot, and there is a continuous evolution connecting them. On a fixed topology run at $Q=0$, some topological activity is present and objects of charges around $\pm 1$ emerge. Early results suggest that the peaks tend to be flatter and broader than on the topological charge changing ensembles, and they emerge by the joining together of several smaller fluctuations; the process is much slower than on the ensembles with topological index changes. We cannot yet say how much slower. 
 
Alternatively, we can use the following measure to compare the local topological charge density of one configuration with another 
\begin{gather}
C(\tau,\tau') =  \frac{\sum_x (q(x,\tau') - \langle q(\tau') \rangle_\tau)(q(x,\tau) - \langle q(\tau)\rangle_\tau}{\sqrt{\left[\sum_x (q(x,\tau') - \langle q(\tau')\rangle_\tau)^2\right]\left[\sum_x (q(x,\tau) - \langle q(\tau)\rangle_\tau)^2\right]}},\label{eq:26}
\end{gather}
where $q(x,\tau)$ represents a definition of the local topological charge density (as before, we use the definition outlined in the appendix). $\langle q(\tau) \rangle_\tau$ represents the average topological charge per lattice site on a given configuration, $\langle q(\tau)\rangle_\tau = \frac{1}{V} \sum_x q(x,\tau)$. This compares the local charge density on two different configurations, normalised to give 1 if they are completely correlated, and zero if they are completely uncorrelated, and with the average charge on the configuration subtracted to reduce any bias if both configurations have a large global charge (without this subtraction, one might expect some correlation to be measured even for uncorrelated configurations). This measure is not perfect, as it does not distinguish between the creation and annihilation of lumps and their movement by a few lattice spacings, but we can expect to see some effect from dramatic changes in the action.  We obtain $C(\tau,\tau+1)=0.39(3)$ for those configurations 
where there is a topological index change, and $C(\tau,\tau+1)=0.72(1)$ for those configurations where there is not. Although our statistics for two charge changes in the same trajectory, one positive and one negative so that there is no overall net effect on the topological index, are small, our results suggest that in this case $C(\tau,\tau+1)$ is usually consistent with the single charge change or smaller. 

It therefore seems likely that allowing topological charge changes will allow a considerably improved sampling of different topological vacuums even between configurations with the same topological charge. We cannot say whether these configurations are accessible for HMC algorithms without topological charge changes within a reasonable time. What is needed is a comparison of autocorrelation time against time per trajectory; but a reliable measurement of the autocorrelation requires considerably more configurations than we have so far.

In the continuum, the topological sectors are disconnected: it is not possible to change the topological charge by continuously deforming the gauge field (or, equivalently, by continuously deforming the Dirac operator). On the lattice, the topological charge is not exactly conserved, allowing us to sneak instantons or anti-instantons through the gaps in the lattice. However, at some small lattice spacing, any lattice Dirac operator will encounter problems while changing topology using an update procedure based on a continuous evolution of the gauge field (or where the gauge field is updated in small steps, as in HMC): if it does not, it does not have the correct continuum limit. This has been observed at around $0.05$fm~\cite{Schaefer:2010hu}. In terms of the overlap topological index, there is a suppression of near-zero eigenvalues of the kernel operator (within the matrix sign function), meaning that there are no eigenvalue crossings of zero, and no topological index changes. A solution to this problem may 
involve either introducing a new Monte Carlo algorithm which gives a large change in the gauge field at each step, or a method of interpolating from gauge fields generated on a coarse lattice to those generated on a fine lattice. Whether such solutions are necessary depends on whether the lack of global topological index changes leads to a poor sampling of gauge fields with clearly distinct local topology.

However, for overlap, or other Ginsparg-Wilson Dirac operators, it is not possible to change the topological index without a discontinuous change in the Dirac operator at any lattice spacing (here a continuous change in the gauge field may lead to a discontinuous change in the Dirac operator). Rather than a suppression of small eigenvalues (or, more likely, as well as this suppression) the key element that prevents topological index changes is the action discontinuity, $\Delta S$, at the topological sector boundary. It is difficult for a Hybrid Monte Carlo simulation to penetrate this discontinuity; and the larger $\Delta S$, the harder it is to change topological sector.

The pseudo-fermion estimate of the determinant~\cite{PseudoFermions} gives an inaccurately large estimate of $\Delta S$ (see~\cite{Egri:2005cx,DeGrand:2004nq,Cundy:2008zc} and below). It is possible that this exacerbates the problem for all actions, but for overlap fermions it can be treated. One factorises the fermion determinant into a continuous part and a discontinuous part, where the discontinuous part can be calculated exactly without recourse to pseudo-fermions. It is therefore possible to design overlap actions where changing topology has different degrees of difficulty. One can also mimic the situation in the continuum limit by artificially suppressing near zero eigenvalues of the kernel operator~\cite{Hashimoto:2006rb}. (One should note that a decrease in the density of kernel eigenvalues is in some respects advantageous for overlap Hybrid Monte Carlo, since it accelerates the cost to generate a trajectory. This is balanced in part by an increase in the density of kernel eigenvalues at larger 
physical volumes. The difficulty in current overlap simulations without topology suppression is not too few near-zero kernel eigenvalues but too many.) This makes overlap HMC an ideal place to test the effects of the suppression of global topological charge changes. The aim would be to run at least two overlap actions on a large enough volume and tuned lattice spacing (not so large that it makes it harder to change topology, nor too small that there is too strong a suppression of topological charge changes for all actions) all generating the same action, but with algorithms that allow frequent topological charge changes, and one with an algorithm with infrequent charge changes. To generate enough statistics would require a large amount of computer time. Overlap fermions would be preferred for this study because of their accurate treatment of topology. 

It is easy to design an overlap HMC algorithm with relatively infrequent topological index changes. The challenge is designing an algorithm where the index changes are as frequent as possible. The development of such algorithms is necessary to prepare for a study on the effect of constant global topological charge on the local charge density and to improve the efficiency of overlap algorithms. In this work, we present our current overlap HMC production algorithm (which, though based on previous work, has not yet been published in full), and compare it against a modification which promises to have more frequent topological index changes though possibly at the cost of a lower metropolis accept/reject rate. 

\section{Actions}\label{sec:algorithm}
For this work we have used a Tadpole improved L\"uscher-Weisz gauge action~\cite{TILW} at $\beta = 8.15$, which corresponds to a lattice spacing of around $0.11$fm, with two flavours of fermions with a quark mass of $\mu = 0.1$ which corresponds to $m_{\pi} \sim 530 MeV$. The lattice volume was $8^332$. One level of over improved stout smearing~\cite{Moran:2008ra,Morningstar:2003gk} at parameters $\rho = 0.1$, $\epsilon = -0.25$ was applied to the gauge links when constructing the Wilson Dirac operator, the molecular dynamics time-step was $\delta\tau = 0.0115$, and the total trajectory length around 2.25. The molecular dynamics used two time-scales~\cite{SW}, one for the gauge fields, and the other sixteen times slower for the fermion fields. We used the Omelyan integrator~\cite{Omelyan,Takaishi:2005tz}. We applied mass preconditioning of the fermionic force~\cite{Hasenbusch:2001ne}, with the masses tuned so that the forces for each pseudo-fermion field were roughly the same. All our ensembles were started 
either from a previously thermalised ensemble from an earlier study (discarding
around 500 length 1 trajectories), or from one of the configurations generated
from a different ensemble from this study (with the first 10 trajectories
discarded for the topology fixed ensembles). Given that the thermalised
distribution ought to be the same for all our ensembles (baring small
corrections from the topology fixing term), we do not expect much thermalisation
is required for each of our ensembles.

\subsection{Topology fixed overlap}\label{sec:fixedtopology}
The JLQCD collaboration have been running large scale dynamical overlap simulations for several years~\cite{Hashimoto:2006rb,Hashimoto:2008fc,Aoki:2008tq}.  They add an additional term to the action to suppress small kernel eigenvalues which avoids having to use the transmission/reflection algorithm to deal with topological index changes. This has the result of accelerating their code considerably, partly due to the lack of a need to spend computer time resolving the eigenvalue crossings, which is a large proportion of the cost for algorithms based on the transmission/reflection method, and partly because the improved condition number of the kernel operator accelerates the computation of the matrix sign function. This has three principle effects: firstly to fix the HMC in one topological sector, which is regarded as a finite volume effect, and can be treated using a known extrapolation formula~\cite{Aoki:2007ka,Brower:2003yx}; and secondly it will possibly affect topological autocorrelation and ergodicity 
within the topological sector; thirdly, there are additional lattice artefacts from the topology fixing ghost term, but these are small enough that they can be accounted for in the standard $a \rightarrow 0$ extrapolation.

We used an action
\begin{gather}
S_J = S_{YM} + \phi_1^{\dagger} \frac{D^+[\mu_1]}{D^+[\mu]} \phi_1 + \phi_2^{\dagger} \frac{1}{D^+[\mu_1]} \phi_2 + \phi_3^{\dagger} \frac{D^+[\mu_1]}{D^+[\mu]} \phi_3+ \phi_4^{\dagger} \frac{1}{D^+[\mu_1]} \phi_4 + \phi_5^\dagger\frac{K^2 + \Delta}{K^2} \phi_5,\label{eq:TFaction}
\end{gather}  
where $D^\pm$ is the single flavour chirally projected Dirac operator~\cite{Bode:1999dd,Cundy:2005mr}
\begin{gather}
D^\pm[\mu] = \sqrt{2(1+\mu^2)} \pm \frac{1-\mu^2}{2\sqrt{2(1+\mu^2)}}(1\pm\gamma_5) \sign(K)(1\pm\gamma_5). 
\end{gather}
$\mu_1$ is tuned so that the fermionic forces are roughly equal. When working in a non-trivial topological sector, we use the projected Dirac operator with no zero modes, and $\phi_i$ are pseudo-fermion fields used to provide a stochastic estimate of the determinant. $D^+ D^- = D^{\dagger}D$ and 
\begin{gather}
\det(D^{\pm}[\mu]) = \det(D[\mu]) \mu^{\pm Q_f}. 
\end{gather}
The remaining $\mu^{\pm Q_f}$ contribution to the determinant can in principle
be included by hand, modifying $\Delta S$ when there is a topological index
change, or using a term such as $\det( 1 + (1-\mu)/(1+\mu)\sign(K))$ in the
action; however in this case this is not necessary because the topological index
does not change. The final term in the action (\ref{eq:TFaction}) gives an
estimate of $\det(K^2/(K^2 + \Delta))$, which suppresses the small eigenvalues
of $K$ but does not contribute as $a \rightarrow 0$ (the small eigenvalues of
$K$ roughly correspond to Eigenvalues of the Dirac operator $D_W$ with a mass
$\sim m/a$).

\subsection{Small kernel mode suppression}\label{sec:ts}
A disadvantage of HMC with topological charge changes is that the density of kernel eigenvalues increases with the physical volume, and thus the cost per trajectory might grow with the square of the volume (the cost per independent configuration need not have this behaviour, as we might expect that the autocorrelation would decrease as there would be more topological charge changes). A large density of kernel eigenvalues would lead to an increase in the computer time as it will be more expensive to approximate the matrix sign function and there will be an increased number of attempts to change the topological index. This should be partially cancelled out by a decrease in the density of eigenvalues caused by a decrease in the lattice spacing, but only on lattice volumes which are too large for current simulations. In practice, it is advisable on larger physical volumes to avoid this difficulty by suppressing, but not eliminating, the small eigenvalues of the Kernel operator. We achieve this by adding to the 
action a term similar to the topology fixing term,
\begin{gather}
\Delta S_{TF} = \phi_{TF}^\dagger \frac{K^2 + \xi \Delta}{K^2 + \Delta} \phi_{TF}.
\end{gather} 
As in the topology fixed action, the effects of this term will disappear in the continuum limit.
The coefficient $\xi$ can be tuned between $0$ and $1$. $\xi =0$ gives the topology fixing term which completely suppresses the low kernel eigenvalues, removing the problem entirely at the cost of forbidding topological index changes. $\xi = 1$ gives the original action which has, for some ensembles, too many small eigenvalues and thus too large a cost per trajectory. By tuning $\xi$ between these values and simultaneously tuning $\Delta$, we can control the density of small kernel eigenvalues, keeping it constant as we increase the lattice volume. In this way, it is to be hoped that the poor volume scaling per trajectory of overlap fermions can be avoided. It is to be hoped that as the continuum limit is approached, this term will not be necessary. Nor was it required on the smaller lattice volumes used in this study. However, on our main $16^332$ ensembles (to be reported in a later publication), this term is required and successfully balances the time per trajectory with the topological tunnelling rate. 
Our numerical studies suggest that there is a smooth transition in the density of low eigenvalues rather than a discontinuous transition as $\xi$ is varied.

In this way, overlap simulations with topological index changes can be performed on larger lattice volumes without a poor volume scaling in the cost per trajectory. In particular, the methods described below are particularly sensitive to the density of small kernel eigenvalues, and thus a potentially poor volume scaling. Using this method, we will be able to avoid this poor volume scaling and thus this possible objection to these methods. 

While we find adding this ghost term to the action unpalatable, methods such as
this remain the only solutions available to the problem of the volume
scaling while we are restricted to relatively large lattice
spacings.\footnote{An alternative formulation of a similar idea is given
in~\cite{Golterman:2007ni}.}

\subsection{Determinant factorised overlap}\label{sec:2.3}
When running a lattice simulation with dynamical overlap fermions, using standard methods (pseudofermions) to estimate the fermion determinant, the discontinuity in the action is far worse than given in equation (\ref{eq:deltaSreal}). The pseudo-fermion estimate of the 2-flavour determinant contributes a term $\bra{\phi} (1/D^{\dagger} D) \ket{\phi}$ to the action, meaning that the discontinuity is
\begin{gather}
\Delta_{Spseudo} = \bra{\phi} \left(\frac{1-\mu^2}{D^{\dagger}{D}}\right)_+ (\gamma_5 \ket{\psi_0}\bra{\psi_0} +  \ket{\psi_0}\bra{\psi_0} \gamma_5)\left(\frac{2 \sign(\lambda_{0})}{D^{\dagger}{D}}\right)_- \ket{\phi}  ,
\end{gather}
where '-' indicates that the overlap operator (or kernel eigenvalue $\lambda$) is calculated with the original sign of the lowest eigenvalue, just before the crossing, and '+' with the opposite sign and $\psi_0$ and $\lambda_0$ are the kernel eigenvector and eigenvalue which crosses zero. This discontinuity with pseudo-fermions firstly is larger than the real discontinuity, and secondly has a poor scaling with the mass $\Delta S \sim 1/\mu^2$. 

Any noisy estimate of the fermion determinant will, on average, have a worse discontinuity in the action than the true determinant's discontinuity. Let us suppose that $\Delta S > 0$ (as it is almost all the time), and that we have some noisy estimate of the action, parametrised by a single variable $\alpha$ and generated according to some distribution $e^{-F(\alpha,U)}$ defined in terms of a positive definite Hermitian function $F(\alpha,U)$ so that, up to a normalisation, 
\begin{gather}
\int_{-\infty}^{\infty} d\alpha e^{-F_{\pm}(\alpha,U)} = e^{-S_{\pm}(U)}, 
\end{gather}
where the $\pm$ indicates the action after or before the crossing. 

We can introduce this noisy estimator into the HMC algorithm, so that the discontinuity in the action is instead $F_+ - F_-$ (we initially make the simplifying assumption that this is a monotonic function of $\alpha$) and the probability of transmission is
\begin{gather}
P_T = \frac{\int_{-\infty}^{\infty} e^{-F_-(\alpha,U)} d\alpha \min(1,e^{F_-(\alpha,U)-F_+(\alpha,U)})}{\int_{-\infty}^{\infty} e^{-F_-(\alpha,U)} d\alpha}.
\end{gather} 
If we suppose that $F_-(\alpha,U) < F_+(\alpha,U)$ for $\alpha$ smaller than some $\alpha_1$, then
\begin{align}
P_T =& \frac{\int_{-\infty}^{\alpha_1} e^{-F_+(\alpha,U)}d\alpha + \int^{\infty}_{\alpha_1} e^{-F_-(\alpha,U)}d\alpha}{\int^{\infty}_{-\infty} e^{-F_-(\alpha,U)}d\alpha}\nonumber\\
=&  \frac{\int_{-\infty}^{\infty} e^{-F_+(\alpha,U)}d\alpha + \int^{\infty}_{\alpha_1} (e^{-F_-(\alpha,U)} - e^{-F_+(\alpha,U)})d\alpha}{\int^{\infty}_{-\infty} e^{-F_-(\alpha,U)}d\alpha}\nonumber\\
=& e^{S_- - S_+} + \frac{\int^{\infty}_{\alpha_1} (e^{-F_-(\alpha,U)} - e^{-F_+(\alpha,U)})d\alpha}{\int^{\infty}_{-\infty} e^{-F_-(\alpha,U)}d\alpha}.
\end{align}
Since $F_+< F_-$ for $\alpha > \alpha_1$, this second term is negative while the first term is $e^{-\Delta S}$. If $F_+-F_-$ is not monotonic in $\alpha$ then we will need to split the action into several regions, then the argument will proceed unaffected. Similarly, the argument can easily be extended for multiple noisy estimators. Therefore using a noisy estimator of the action, such as the pseudo-fermion estimate of the determinant, can never improve the rate of transmission, and will frequently reduce it considerably. If the topological auto-correlation depends on the rate of topological charge changes, then we need to do better than use a pseudo-fermion estimate of the determinant.

In~\cite{Cundy:2007la,Cundy:2008zc}, it was proposed to factorise the determinant using an approximate sign function $\tilde{\epsilon}(K)$, which is continuous at $K=0$ and differs from the matrix sign function for $n$ eigenvectors with eigenvalues close to zero. The single flavour determinant is factorised according to
\begin{multline}
\det(1+\mu + (1-\mu) \gamma_5 \sign(K)) = \\
\det(1+\mu + (1-\mu) \gamma_5 \tilde{\epsilon}(K))\det\left(1 + \frac{1-\mu}{1+\mu + (1-\mu) \gamma_5 \tilde{\epsilon}(K)}\gamma_5(\sign(K) - \tilde{\epsilon}(K))\right)
\end{multline}
If $\sign(K) - \tilde{\epsilon}(K) = \sum_i c(\lambda_i) \ket{\psi_i}\bra{\psi_i}$ for eigenvectors $\ket{\psi_i}$ and eigenvalues $\lambda_i$ of $K$, then the determinant can be written as
\begin{gather}
\det(D) = \det(\tilde{D}) e^{\Tr \log (\delta_{ij} + \bra{\psi_i} \frac{1}{\gamma_5 \tilde{D}} \ket{\psi_j} c(\lambda_j))},
\end{gather}
 where
\begin{gather}
\tilde{D} = 1+\mu + (1-\mu) \gamma_5 \tilde{\epsilon}(K).
\end{gather}
The determinant of $\tilde{D}$ can be computed using pseudo-fermions, while the
$\Tr\log$ can be computed exactly as it is only an $n\times n$ matrix.
In~\cite{Cundy:2007la}, $\tilde{\epsilon}(K)$ was a Zolotarev approximation to
the matrix sign function. This is continuous, and thus avoids the additional
reflections required by the choice recommended here, but the Zolotarev
approximation must be quite broad (i.e. if the rational approximation is
accurate for eigenvalues $|\lambda | > \xi$, then $\xi$ must by relatively
large) to avoid unacceptably large variation in the fermionic forces when there
is a small eigenvalue of $K$, as the force depends on $d
\tilde{\epsilon}/d\lambda$. The principle reason that we stopped using this
algorithm in the dynamical overlap simulations and switched to the algorithm
described here is that the algorithm with the Zolotarev approximation proved to
be unstable due to the large fluctuations in the fermionic force. Using a
shifted matrix sign function,
\begin{gather}
\tilde{D} = D_n = 1+\mu + (1-\mu) \gamma_5 \sign(K - \Lambda_n),
\end{gather}
has the advantages of that approach
over the pseudo-fermion approach, described in the earlier work, without this
large disadvantage. It is also faster as fewer eigenvectors needed to be
included in the additional determinant, which is more than enough to compensate
for the cost due to the additional discontinuity in the action.
$\Lambda_n$ is chosen at the start of each trajectory within the region
$[-\beta,-\alpha] \cup [\alpha,\beta]$ for coefficients $\beta >\alpha > 0$.
$\alpha$, $\beta$ and the probability distribution  (we just used a flat
distribution) can be tuned to optimize the computer time required for a
trajectory. Choosing $\Lambda_n$ does not affect the stationary distribution of
the Markov chain: once we have integrated out the pseudo-fermions, the transfer
matrix of the Markov process is independent of $\Lambda$. Allowing $\Lambda$ to
change sign ensures that there are no difficulties with ergodicity. $n$
eigenvalues of $K$ lie between $0$ and $\Lambda_n$: in effect we have modified
the matrix sign function so that these $n$ eigenvalues have flipped
sign\footnote{We use this notation for later convenience, although it can be
deceptive: in practice $\Lambda_n$ is chosen randomly, then $n$ can be counted.
We do not in the simulations select $n$ and then choose a $\Lambda_n$
accordingly. $\Lambda_n$ is not a direct function of $n$; rather $n$ is 
determined by the previous choice of $\Lambda_n$. In our simulations, $n$
fluctuated between $3-5$}. The ratio of the determinants between this Dirac
operator and the overlap Dirac operator before the eigenvalue crossing (where $D
 = D_0$) is
\begin{gather}
\frac{\det{D_n}}{\det D_0} = \frac{\det{D_1}}{\det D_0} \frac{\det{D_2}}{\det D_1} \frac{\det{D_3}}{\det D_2}\ldots \frac{\det{D_n}}{\det D_{n-1}}.
\end{gather}
These Dirac operators differ by having different number of eigenvalues flip sign in relation to $D_n$. After the eigenvalue crossing, the lowest eigenvalue of $D$ flips sign, so we will now have either $D=D_1$ or $D=D_{-1}$ depending on whether an instanton is created or destroyed; to be definite let us suppose that it becomes $D_1$. The change in the HMC action is then
\begin{gather}
\Delta S = \Tr \log\left[\frac{\det{D_1}}{D_n}\right] -  \Tr \log\left[\frac{\det{D_0}}{\det D_n}\right] = \Tr \log\left[\frac{\det{D_1}}{\det D_0}\right] = \Delta S_{real}.
\end{gather}  
Therefore, this particular choice of determinant factorisation preserves the
best possible transmission rate for an exact, energy conserving, algorithm
(indeed any continuous $\tilde{D}$ where $\det \tilde{D}/D$ can be easily
calculated could be used and will give the correct $\Delta S$). The
corresponding HMC algorithm is as stable as the original overlap HMC
algorithm~\cite{Cundy:2007df}, and allows another advantage of overlap HMC:
chiral factorisation to permit single flavour
simulations~\cite{Bode:1999dd,Cundy:2005mr} without RHMC. Because
$\tilde{\epsilon}$ differs from epsilon for only eigenvalues between $0 <
\lambda < \Lambda_n$ rather than $-\xi < \lambda < \xi$ for a continuous
approximation to the matrix sign function, the correction determinant is smaller
and thus cheaper to calculate. By using the low kernel mode suppression
described in section \ref{sec:ts} if necessary, we can ensure that $\Lambda_n$
is large enough that eigenvalues do not bounce between the two discontinuities
(which increases the cost either by having a large number of reflections or
because the topological index change is rapidly undone by the eigenvector
crossing zero) while keeping 
$n$ small enough that the cost of calculating the small determinant is under control. Using this $\tilde{\epsilon}$ comes, however, at the cost of an additional reflection when an eigenvalue of $K$ attempts to pass $\Lambda_n$. There is no point in attempting to transmit at this new barrier because $\Delta S$ will be too large.

The action used in this study is
\begin{multline}
S_F = S_{YM} + \phi_1^{\dagger} \frac{D_n^+[\mu_1]}{D_n^+[\mu]} \phi_1 +
\phi_2^{\dagger} \frac{1}{D_n^+[\mu_1]} \phi_2 +  \phi_3^{\dagger}
\frac{D_n^+[\mu_1]}{D_n^+[\mu]} \phi_3 +   \phi_4^{\dagger}
\frac{1}{D_n^+[\mu_1]} \phi_4 -\\ {n_f \Tr \log \left(\delta_{ij} + \bra{\psi_i}
\frac{1}{\gamma_5D_n} \ket{\psi_j}\left(\sign(\lambda_j) - \sign(\lambda_j -
\Lambda_n)\right)\right)},\label{eq:23} 
\end{multline}
where $n_f = 2$ is the number of flavours, and $0\le i,j <n-1$.

We have successfully deployed this action in production runs for lattices of volumes up to $16^332$ since shortly after the publication of~\cite{Cundy:2007la}.

\subsection{Inexact Determinant Factorised}\label{sec:idf}
Up to finite volume corrections, and assuming a Gaussian momentum distribution, the probability of accepting a topological index change as a kernel eigenvalue crosses zero is $\min(1,e^{-\Delta S})$~\cite{Cundy:2005mr}. The probability of accepting the trajectory in the final Monte-Carlo step were the transmission/reflection algorithm neglected would be $\min(1,e^{-\Delta S})$, if the molecular dynamics time step, $\delta\tau$, is small enough that the errors in action conservation during the molecular dynamics are negligible. If we were only interested in the rate of topological index changes and nothing else (which, of course, we are not) then there would be no difference in the efficiency of the algorithm whether we just neglected the transmission/reflection procedure and dealt with the action jump in the molecular dynamics step or used the transmission/reflection algorithm to `exactly' (up to O($\delta\tau^3$) corrections) conserve energy. Indeed, we would gain by omitting the transmission/reflection 
algorithm because of the cost of this part of the code. One can also in principle distribute the burden of changing the topological index between the metropolis step and the transmission/reflection routine; partly compensating for $\Delta S$ in one, and partly in the other. Of course, we do not only wish to sample each topological sector but sample within the topological sector, and for this reason it is better to use the transmission/reflection algorithm and reject only the topological index change rather than reject the entire trajectory. 

Suppose that we have a different action $S'$ conserved by the molecular dynamics to the action $S$ which defines the distribution and is included in the metropolis step (in practice, this is always true: the molecular dynamics exactly conserves an action that differs from $S$ by terms of order $\delta\tau^2$~\cite{Clark:2007cb}). In this case, there would be a lower acceptance rate. However, if the molecular dynamics for the action $S'$ better samples the phase space than the molecular dynamics constructed to conserve $S$ and the difference $S' - S$ is not large and scales well with the lattice volume (which for most choices of $S'$ it doesn't) it may be beneficial to trade acceptance rate for a shorter auto-correlation time for those trajectories which are accepted.  

We here propose using within the molecular dynamics, but not the metropolis step (where the exact overlap action (\ref{eq:23}) is used), an action modified by a term designed to partially cancel out the action discontinuity. The proposed modification to the action is 
\begin{align}
S'_I =& S_F + S_I\nonumber\\
S_I = &\sum_i \frac{\alpha}{2\mu_i}n_{fi}  \log\left(1+\bra{\psi_i} \frac{2 g(\lambda_i)\sign(\lambda_i) \mu_i}{\gamma_5 D[\mu_i]} \ket{\psi_i}\right),\label{eq:24}
\end{align}
$\alpha$ is a tunable parameter, $n_{fi}$ the number of flavours with mass $\mu_i$, $g(\lambda)$ is a function which satisfies $g(\delta)\sign(\delta) - g(-\delta)\sign(-\delta) = 2$ for $\delta \rightarrow 0$ and decays to zero as the eigenvalue moves away from 0. For our tests, we have used $g(\lambda) = 2 (1- (\lambda/\Lambda_n)^2)^3$ for $\lambda$ between 0 and $\Lambda_n$, and zero otherwise. To avoid discontinuities in the action, we need to sum over all the eigenvectors and not just the lowest one. Because, using the suppression of \ref{sec:ts} if required, the number of eigenvalues contributing to $S_I$ is held constant as the physical volume increases and the expression for $S_I$ is intensive we do not expect in this case that using a different action in the molecular dynamics and metropolis parts of the algorithm will lead to a poor volume scaling.

If we added the quantity $\log(1+\bra{\psi_0} \frac{ g(\lambda_0)\sign(\lambda_0)}{\gamma_5 D} \ket{\psi_0})$ to the action, the discontinuity in this operator would exactly cancel the discontinuity in the fermion determinant, allowing a 100\% transmission rate. This in practice is not possible, because of instabilities as $\bra{\psi_i} \frac{ g(\lambda_i)\sign(\lambda_i)}{\gamma_5 D} \ket{\psi_i} \sim -1$ (and the transmission rate is also affected by discontinuities in the terms from the other eigenvectors), so we have modified the action as above. At small $\bra{\psi_0} \frac{1}{\gamma_5 D} \ket{\psi_0} \ll 1$, this would exactly cancel the discontinuity; at larger $\bra{\psi_0} \frac{1}{\gamma_5 D} \ket{\psi_0}$ it tends to over-estimate the action discontinuity. Were $g$ constant (and there only one eigenvalue between 0 and $\Lambda_n$), this procedure would, for a correctly tuned $\alpha $, transfer the possibility of rejection when changing the topological index from the transmission routine to the 
metropolis step.  

But in practice $g$ is chosen so that as the eigenvalues move away from 0 the additional term in the action becomes small. The action conserved by the molecular dynamics approaches the action within the metropolis step. This means that the metropolis acceptance rate may remain high while still allowing an enhanced number of topological index changes: unless the trajectory starts or ends with an eigenvalue close to $0$, the error in the action for the metropolis step, $S_I$, will be small. For this reason, only the smallest kernel eigenvalue will give a large contribution to $S_I$. Of course, this requires the force from this term is stable (which means that $g$ cannot be too sharply peaked), and, to avoid poor volume scaling, one would need to suppress, but not eliminate the small kernel eigenvalues on larger volumes (possibly using the method of section \ref{sec:ts}) to ensure that $\Lambda_n$ is as large as possible while still having only a small number of kernel eigenvectors in the range $0<\lambda/\
Lambda_n < 1$. If this is done, then $S_I$ will not grow with the lattice
volume, and could be used without additional cost on larger lattices.
Additionally, we may tune the coefficient $\alpha$: $\alpha = 0$ gives the
original action with a good metropolis acceptance rate but poor transmission
rate; $\alpha = 1$ gives an action with (presumably) a better transmission rate
and poorer metropolis acceptance rate. By tuning $\alpha$, we can hope to find
an optimal balance between acceptance rate and transmission rate -- there is
$\textit{a priori}$ reason to suspect that the optimum is at $\alpha = 0$ or
$\alpha = 1$. One feature of this method is that a once an eigenvalue
successfully crosses zero, there is a strong force invariably trying to push it
back; topological index changes occur, but often change back within the course
of a trajectory.

$S_I$ can be included in the Monte Carlo at negligible additional cost. During the calculation of $S_F$, we have already solved for $(D_n)^{-1} \ket{\psi_i}$ for all the eigenvectors of interest. Noting that
\begin{gather}
D_{n-1} = D_n + 2 \sign(\lambda_{n-1}) \ket{\psi_{n-1}}\bra{\psi_{n-1}},
\end{gather}    
and by using the Sherman-Morrison formula~\cite{ShermanMorrison}
\begin{gather}
\frac{1}{M + \ket{A}\bra{B}}\ket{C} = \frac{1}{M}\ket{C} - \frac{\frac{1}{M}\ket{A}\bra{B}\frac{1}{M}}{1+\bra{B}\frac{1}{M} \ket{A}}\ket{C}
\end{gather}
it is easy to construct the solutions for $D_{n-1}^{-1}\psi_i$, and, by a simple process of iteration, obtain $D_0^{-1}\psi_i$.

For our simulations, we used $\alpha = 0.75$ which gave a HMC acceptance rate of
$80\%$ and a transmission rate of $70\%$. We did not have the computing
resources to devote to accurately tuning $\alpha$ and $g$ (beyond an
initial test on $4^38$ lattices), and there is no reason to suppose that our
choices are optimal.

\section{Results}\label{sec:results}
\subsection{Ensembles}
We generated ensembles using the different overlap HMC algorithms for two flavours of fermions on $8^3\times32$ lattices:
\begin{enumerate}
\item \textbf{Ensemble 1}: This used the action described in equation (\ref{eq:23}) which represented the state of the art method for overlap fermions (with topological charge changes) before this work;
\item \textbf{Ensemble 2}: This used the new proposed action described in equation (\ref{eq:24}) for the molecular dynamics and the action of equation (\ref{eq:23}) for the metropolis step. It is therefore distributed according to the same action as Ensemble 1;
\item \textbf{Ensemble TF1}: This used the topology fixed action incorporating the term of equation (\ref{eq:TFaction}) into the action of (\ref{eq:23}) with $\Delta = 0.2$, with topological charge held constant at (fermionic charge) $Q=-3$;
\item \textbf{Ensemble TF2}: This used the topology fixed action incorporating the term of equation (\ref{eq:TFaction}) into the action of (\ref{eq:23}) with $\Delta =0.2$, with topological charge held constant at (fermionic charge) $Q=0$.
\end{enumerate}
 On each ensemble, we generated over 100 length 2.25 trajectories.  All the ensembles were generated with the same molecular dynamics time-steps and Omelyan and Hasenbusch parameters. Apart from the topological fixing term, which would give a small lattice artefact, all our ensembles describe exactly the same action: the only differences are the different algorithms used, and that two ensembles were in a topological fixed sector. The ensembles were generated on an 8 core 2.93GHz Intel processor on a desktop computer. The details of the ensembles are given in table \ref{tab:1}.
\begin{table}
\begin{tabular}{l l l l l l l l l l}
Ensemble&$\beta$&$\mu$&$r_0/a$&$m_\pi a$&$P_{acc}$&$P_T$&$n_T$&$n_{traj}$&time (s)\\
1&$8.15$ &0.1 &4.64(5)&0.387(14) & 84.1\%&29.9\%&38&151&48688
\\
2&$8.15$ & 0.1&4.66(5)&0.394(9)& 82.1\%&72.2\%&117&162&54026
\\
TF1 &$8.15$ & 0.1& 4.58(4)& 0.392(12)&86.1\%&-&-&101&17801
\\
TF2 &$8.15$ & 0.1& 4.52(4)&0.362(13)&86.8\%&-&-&167&15699
\end{tabular}
\caption{$\beta$, $\mu$, $r_0$ and pion mass $m_\pi$ in units of the lattice spacing, HMC acceptance rate, transmission rate, number of topological index changes, number of length 2 trajectories and time in seconds per trajectory for each of our ensembles. $r_0\sim 0.49\text{fm}$ was measured using the potential extracted from Wilson loops~\cite{Sommer:1993ce}.}\label{tab:1}
\end{table}
 The topology fixed ensemble at $Q=0$ has a slightly larger lattice spacing and smaller pion mass, but we do not expect this to be significant. This is likely to be caused by a $\beta$ shift and finite volume effect from the topology fixing and the additional term in the action. Nonetheless, the difference is not large enough to expect a significant effect on the QCD vacuum. The cost per trajectory for the HMC algorithms with topological charge changes is roughly a factor of 3 slower than the fixed topology ensembles. The greater cost per trajectory for ensemble 2 compared to ensemble 1 is mainly caused by an increase in the number of attempted topological charge changes and reflections of the surface of $\Lambda_n$ eigenvalues. The HMC acceptance rate is good and consistent for all our ensembles. We do not see any significant diminishing of the acceptance rate caused by the re-weighting term in ensemble 2. However, there are over twice as many topological charge changes for ensemble 2 compared to ensemble 1 
(although many of them would have been reversed in the same trajectory for both ensembles, and thus will not appear in figure \ref{fig:1}). The new term proposed in this paper therefore performs as we had hoped: it improves the rate of index changes without adversely affecting the acceptance rate.  
\subsection{Topological index changes}
\begin{figure}
\includegraphics[width=0.95\textwidth]{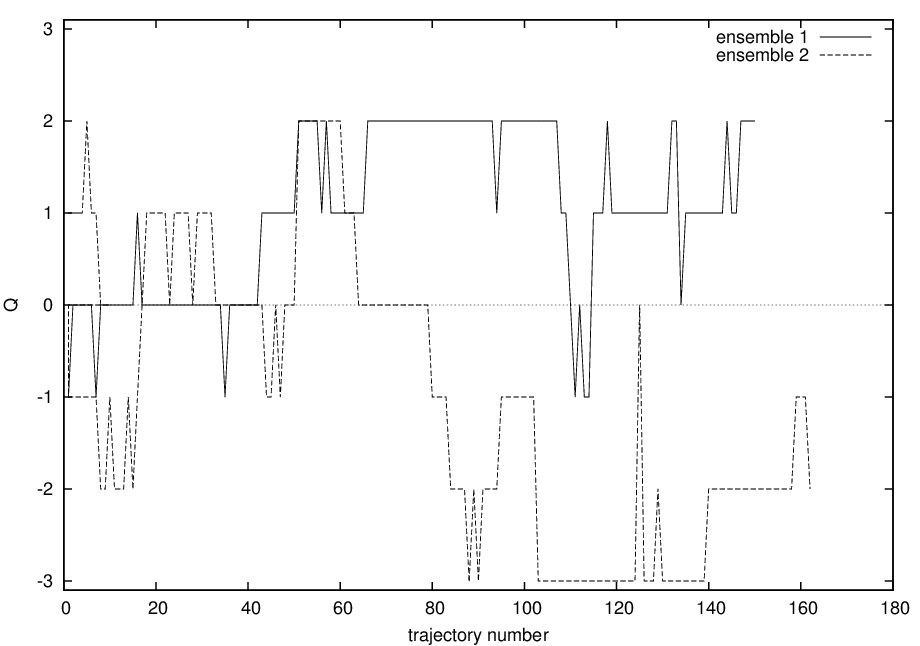}
\caption{Topological index histories for ensembles 1 and 2 (measured using the overlap definition of the topological index).}\label{fig:1}
\end{figure}

The topological index history for ensembles 1 and 2 is shown in figure \ref{fig:1}. Both sample different topological sectors, but ensemble 2 samples the sectors more rapidly. 

The rate of topological index changes is still not large for ensemble 2 compared to other runs we have performed at larger volumes: the rate of index changes is suppressed by the small lattice volume. Our ensembles are large enough to compare rates of topological index changes, but clearly any estimate of the autocorrelation of the global topological charge is impossible.

\subsection{Distribution of $\Delta S$} 
\begin{figure}
\begin{center}
\begin{tabular}{c}
\includegraphics[width = 0.85\textwidth]{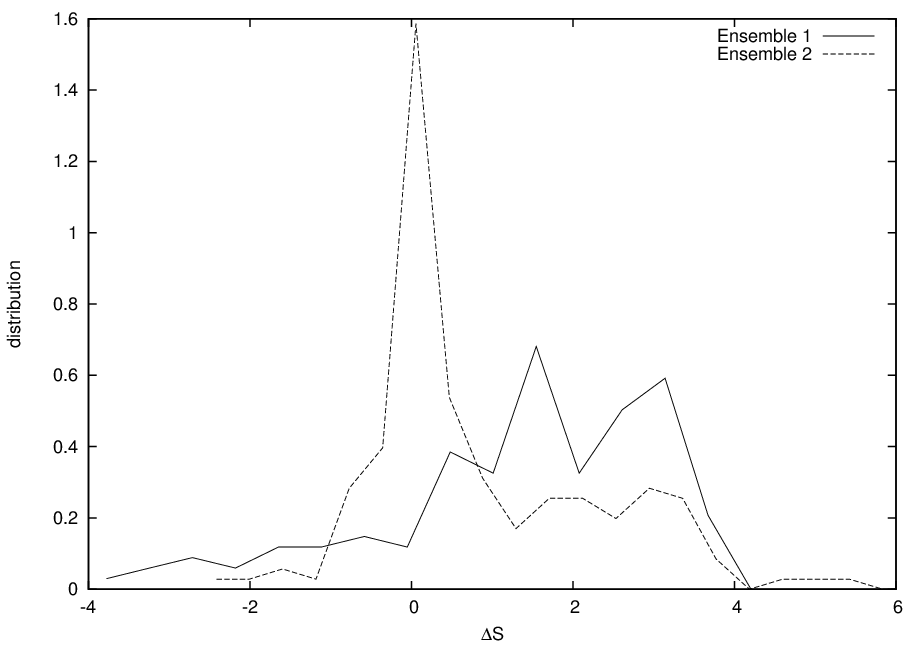}
\end{tabular}
\end{center}
\caption{The distribution of the action discontinuity $\Delta S$ for ensembles 1 and 2.}\label{fig:DeltaS}
\end{figure}

The distribution of $\Delta S$, the discontinuity in the action across the  for
ensembles is shown in figure \ref{fig:DeltaS}. Each attempted topological index
change had a particular $\Delta S$; the plot shows the distribution of these
discontinuities. The probability of a topological index change is
(approximately) $\min(1,e^{-\Delta S})$, so any $\Delta S \lesssim 1$ has a good
chance of allowing a topological index change. For ensemble 1, $\Delta S$ is
reasonably small, with a broad peak between $0<\Delta S<4$. (A corresponding
plot for pseudofermions with mass preconditioning was given
in~\cite{Cundy:2008zc}, and the peak extends far off the right hand side of this
plot. Additionally, poor mass scaling means that it is virtually impossible to
change the topological index on realistic ensembles using only a pseudo-fermion
estimate of the determinant). For ensemble 2, there is a sharp peak around
$\Delta S = 0$, indicating that the additional term in the action correctly
cancelled the action discontinuity in most cases.
 There is a smaller number of occasions where $\Delta S = 0$ remains large. These are caused either by the contributions of eigenvectors other than the one which crossed zero to the action in equation (\ref{eq:24}), or if $\bra{\psi_i} \frac{2 \sign(\lambda_i)}{\gamma_5 D[\mu_i]} \ket{\psi_i}$ is particularly large so that the approximation used to derive this expression breaks down. Nonetheless, for the most part the additional term in the action correctly cancelled out the discontinuity, allowing the higher rate of topological index changes.

\section{Conclusions}\label{sec:conclusions}

The aim of this study was to test whether it is possible to increase the rate of topological charge changes for dynamical overlap fermions by adding to the molecular dynamics an additional term designed to partially cancel out the discontinuity in the action. The danger is that this may lead to a poor metropolis acceptance rate. However, with very little tuning, we were able to achieve a case where the acceptance rate was left unchanged, but the distribution of $\Delta S$ and thus the rate of topological index changes improved considerably. We note that the original algorithm had the best possible rate of topological charge changes for algorithms where the metropolis action is conserved in the molecular dynamics. We do not see a significant increase in the cost per trajectory for the new algorithm.

We have also described a method by which the poor volume scaling in the cost per
trajectory for overlap fermions may be avoided. This is achieved by adding a
term to the action suppressing the small kernel eigenvalues. This will add
additional artefacts, but, until a better method is found, seems to be the best
option available.

Two important questions unaddressed in this study and left for future work are
the volume scaling of the new algorithm (in particular whether the HMC
acceptance rate scales poorly with the volume) and autocorrelation. We do not
expect a problem with the volume scaling, since the addition to the action is an
intensive quantity, but this needs to be confirmed numerically. The issue of
autocorrelation is crucial, in particular to determine if the additional cost
per trajectory for allowing topological charge changes is sufficient to beat the
topology fixed simulations, and to answer questions which are arising for other
actions as the continuum limit is approached. However, we require (and are
generating) more statistics before we are able to address this.

\section*{Acknowledgements}
Computer simulations were carried out on a desktop computer provided by Seoul National University.
Funding was provided by the BK21 program of the NRF, Republic of Korea. The
research of W. Lee is supported by the Creative Research Initiatives program
(3348-20090015) of the NRF grant funded by the Korean government (MEST). W. Lee
would like to acknowledge the support from KISTI supercomputing center through
the strategic support program for the supercomputing application research [No.
KSC-2011-C3-03]. Some of work in this article was supported by the SFB TR-55
funded by the Deutsche Forschungsgemeinschaft.

\appendix\section{5 Link topological charge}
Because of limited computer resources, we used a gluonic definition of the topological charge density, using a 5 link improved topological charge operator~\cite{deForcrand:1997sq}, with gentle over improved stout smearing~\cite{Moran:2008ra,Morningstar:2003gk} ($\rho = 0.03, \epsilon = -0.25$) to remove ultra violet fluctuations. Although the fermionic overlap definition of the local charge density~\cite{Horvath:2003yj,Hasenfratz:1998ri} would ideally be preferred, it is considerably more expensive (requiring a computation of numerous overlap eigenvalues to a high accuracy). Studies have shown that there is little difference between different definitions of the local charge density~\cite{Cundy:2006PoS,Solbrig:2007nr} on most lattices. We applied up to 50 smearing sweeps, and, except when displaying the effect of different levels of smearing, all measurements are after 50 sweeps. In almost every case, the topological charge converged to an integer $\pm \sim 0.05$ after 20 sweeps and an integer $\pm \sim 0.02$ 
after 50 sweeps. The gluonic global topological charge agreed with the fermionic topological index (measured by counting the zero modes of the Dirac operator) over 90\% of the time; for example 10 out of 106 configurations in ensemble 2 had a mismatch in one unit of charge. The discrepancies were related to bumps in the plot of the topological charge against smearing sweeps, suggesting that some of the global charge was destroyed by the smearing. The $Q=3$ topology fixed distribution had 2 discrepancies between the fermionic and gluonic definitions of the charge, while the $Q=0$ topology fixed distribution had no discrepancies between the overlap and gluonic charges.

\begin{figure}
\includegraphics[width = 0.95\textwidth]{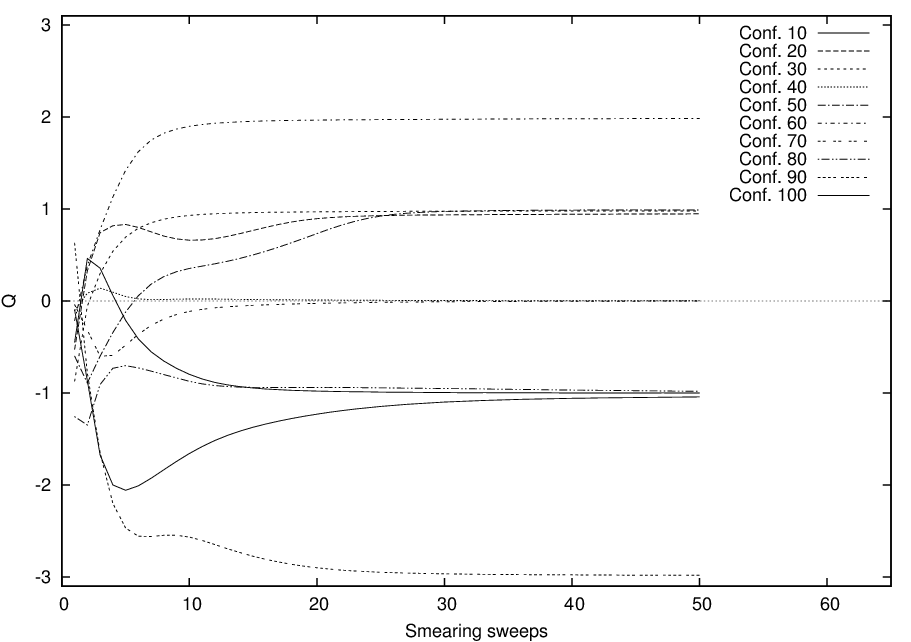}
\caption{The global topological charge for each sweep of over improved smearing for various configurations of ensemble 2.} \label{fig:2}
\end{figure}

The history of the gluonic charge for each smearing sweep for configurations from ensemble 2 is shown in figure \ref{fig:2}. It can be seen that the global charge converges well to an integer value after around 30 smearing sweeps.
\bibliographystyle{elsarticle-num}
\bibliography{weyl}
\end{document}